# *Two step solid state synthesis and Synchrotron X-ray characterizations of ceramic $Co_3TeO_6$; an improper multiferroic**

**Harishchandra Singh,[a,b♣] A. K. Sinha,[b] S. M. Gupta,[c] M. N. Singh,[b] H. Ghosh[b]**

[a]Homi Bhabha National Institute.
[b]Indus Synchrotrons Utilization Division.
[c]Laser Materials Development and Devices Division.
Raja Ramanna Centre for Advanced Technology, Indore – 452013, India.

## Abstract

A two step solid state reaction route has been presented to synthesize monophasic cobalt tellurate ($Co_3TeO_6$, CTO) using $Co_3O_4$ and $TeO_2$ as starting reagents. During synthesis, initial ingredient $Co_3O_4$ is found better than CoO in circumventing the intermediate $Co_5TeO_8$ or $CoTeO_3$ phases. High resolution Synchrotron X-ray Diffraction has been used to probe different phases present in synthesized CTO and to achieve it's single phase. Further, XANES studies near Co K and Te L-edge reveal mixed oxidation states of Co (i.e. $Co^{2+}$ and $Co^{3+}$) and +VI valence state of Te respectively, which is also confirmed with XPS. Charge imbalance due to different oxidation states of the Co-ions has been observed to be compensated by plausible Te-cations vacancy. Enhanced multiferroic properties like effective magnetic moment (JAP **116**, (2014)) have been correlated with the present synthesis route.

## 1. Introduction

Multiferroics where coupling phenomena among various ferroic order parameters e.g. mutual control of polarization (P) by magnetization (M) have attracted great attention due to their promising potential application as spintronics and magnetic data storage devices, and recently in solar cell [1-3]. In the quest of new and better performing materials, cobalt tellurate $Co_3TeO_6$ (CTO) seems to have favorable characteristics [4-7]. The renewed interest in CTO is due to its low temperature type II multiferroic property. Complexity in CTO mainly emerges due

♣ Corresponding author. Tel.: +91-731-244 2583; Fax: +91-731-244 2140.

E-mail address: singh85harish@gmail.com, singh85harish@rrcat.gov.in

to monoclinic (C*2/c*) structure where five different crystallographic sites of Co ions are coordinated by {4}, {5} and {6} oxygen atoms which show subsequent incommensurate antiferromagnetic (AFM) transitions [4-7]. In addition to coupled ferroics orders, a number of multiferroics possessing cationic or anionic non-stoichiometry are useful in many other applications, such as energy conversion, oxygen sensing and oxygen storage [8,9]. Non-stoichiometry (of cations / anions) depends on preparation condition as material synthesized under different conditions behaves differently. For example, materials like $LuFe_2O_{4+\delta}$, $YBaCo_4O_{7+\delta}$ etc. synthesized using same solid state reactions but under different conditions show oxygen storage capacity and structure flexibility, in addition to their intrinsic behaviors.[9] In the similar framework, we synthesize $Co_3TeO_6$ via conventional solid state reaction route using cobalt oxide ($Co_3O_4$) and tellurium oxide ($TeO_2$) as the initial ingredients**.** Origin and the complete understanding of complex magnetism in a monophasic CTO remains a challenging issue [7,10-15].For example, whether the magnetic transition at ~ 34 K observed both in single crystal as well as ceramic CTO corresponds to intrinsic CTO or to impurity, is demonstrated here.

There are mainly two ways; hydrothermal or solid phase reaction by which the CTO synthesis has been reported. Different starting reagents like CoO, $Co(NO_3)_2.H_2O$, $TeO_2$, $Te(OH)_6$, $H_2TeO_4.2H_2O$ and tellurium oxide ($TeO_3$) have been used to synthesize CTO. Thermal decomposition of cobalt (II) tellurate-molybdates ($CoTeMoO_6$ and $Co_4TeMo_3O_{16}$) and amorphous cobalt tellurate $(CoH_4TeO_6.Co(OH))_2$ around 900 K for 24 hrs in air has also been reported for the CTO synthesis [13]. Different second phases like the $Co_5TeO_8$ and $CoTeO_3$ are generally present in the calcined powder and in ceramics, which has been related to loss of $TeO_2$ oxide during calcination/sintering at high temperature [10,12].Thermographic study of the CoO-$TeO_2$ system heated in air has revealed no oxidation to tellurate, unlikely to that observed in NiO-$TeO_2$ system [11]. Synthesis of the CTO single crystal has been reported by Beckar et al.[14] using 3:2:1 ratio of $Co_3O_4$ + $TeO_2$ + $CoCl_2$ (or $PtCl_2$) in the transport reaction method.

It is known that the CTO decomposes spontaneously above 1100 K to CoO and $TeO_2$ but direct synthesis of the single phase CTO from the CoO and $TeO_2$ below 1100 K seems difficult [10].Till now, no reaction mechanism has been proposed for the CTO formation. Here we report, synthesis of the single phase CTO using two step reactions between $Co_3O_4$ and $TeO_2$ in air at <

1100 K. The sample has been found to show better multi-ferroic properties [5,6] which are due to the presence of mixed oxidation states of Co-ions, specially the high spin state of $Co^{3+}$.

## 2. Experimental: Growth and Instrumentation

Cobalt tellurate ($Co_3TeO_6$) is synthesized using conventional solid state reaction route. The reactants used are of analytical grade i.e. cobalt oxide $Co_3O_4$ (Alpha Easer 99.7 %) and tellurium dioxide $TeO_2$ (Alpha Easer 99.99 %). First the reactants are taken in proper stoichiometric ratio and mixed thoroughly using Shaker mixture (TURBULA @- T 2 F) for 6 hrs. Further, the same procedure has also been done with non- stoichiometric ratio of $Co_3O_4$ and $TeO_2$. Obtained samples have been characterized thoroughly in order to optimize the synthesis procedure.

Spectroscopic and structural studies have been done using theSXRD, XANES and XPS measurements at INDUS Synchrotron sources [16]. SXRD measurements are performed at angle dispersive X-ray diffraction (ADXRD) beam-line (BL -12) [17].The beamline consists of a Si ( 111) based double crystal monochromator and two experimental stations namely a six circle diffractometer with a scintillation point detector and Mar 345 dtb image plate area detector. SXRD measurements are carried out using both six circle diffractometer and the image plate. The X-ray wave length used for the present study was accurately calibrated by doing X- ray diffraction on $LaB_6$ NIST standard. Data reductions (taken at image plate) are done using Fit2D software [17]. XANES measurements (performed at BL -12) are carried out in transmission mode (for Co K-edge) at room temperature. The photon energy are calibrated by the Co K-edge XANES spectra of standard Co metal foils at 7709 eV. Te $L_3$-edge XANES spectra has been recorded at Scanning EXAFS Beamline (BL-9) [18]. XPS measurements are performed at room temperature with an OMICRON 180° hemispherical analyzer (model EA 125) using Al Kα (photon energy = 1486.7 eV) source [19]. The hemispherical analyzer is operated in constant-pass energy mode and its overall energy resolution with pass energy of 50 eV is estimated to be 0.8 eV. The measurements are carried out with a photoelectron take-off angle of 45° and the pressure in the spectrometer chamber during the measurements was $\sim 10^{-10}$ mbar [19]. Magnetic measurements are carried out with a vibrating sample magnetometer in the magnetic property measurement system (MPMS-SQUID VSM) of Quantum Design, USA. The temperature stability at the sample chamber is better than ± 0.2 K. Thermal analysis, such as differential thermal analysis (DTA) and thermo-gravimetric analysis (TGA) of the oven-dried powder is carried out in a thermal analyzer (Perkin Elmer, Pyris Diamond) at a heating rate of 10$^o$C/min. Thermal analysis of the mixed powder is carried out to determine the reaction mechanism.

## 3. Results and Discussion

Fig. 1 presents the thermo-gravimetric (TG), differential thermal analysis (DTA) and differential thermal gravimetric (DTG) plots of the $Co_3O_4$+$TeO_2$ mixed powder.

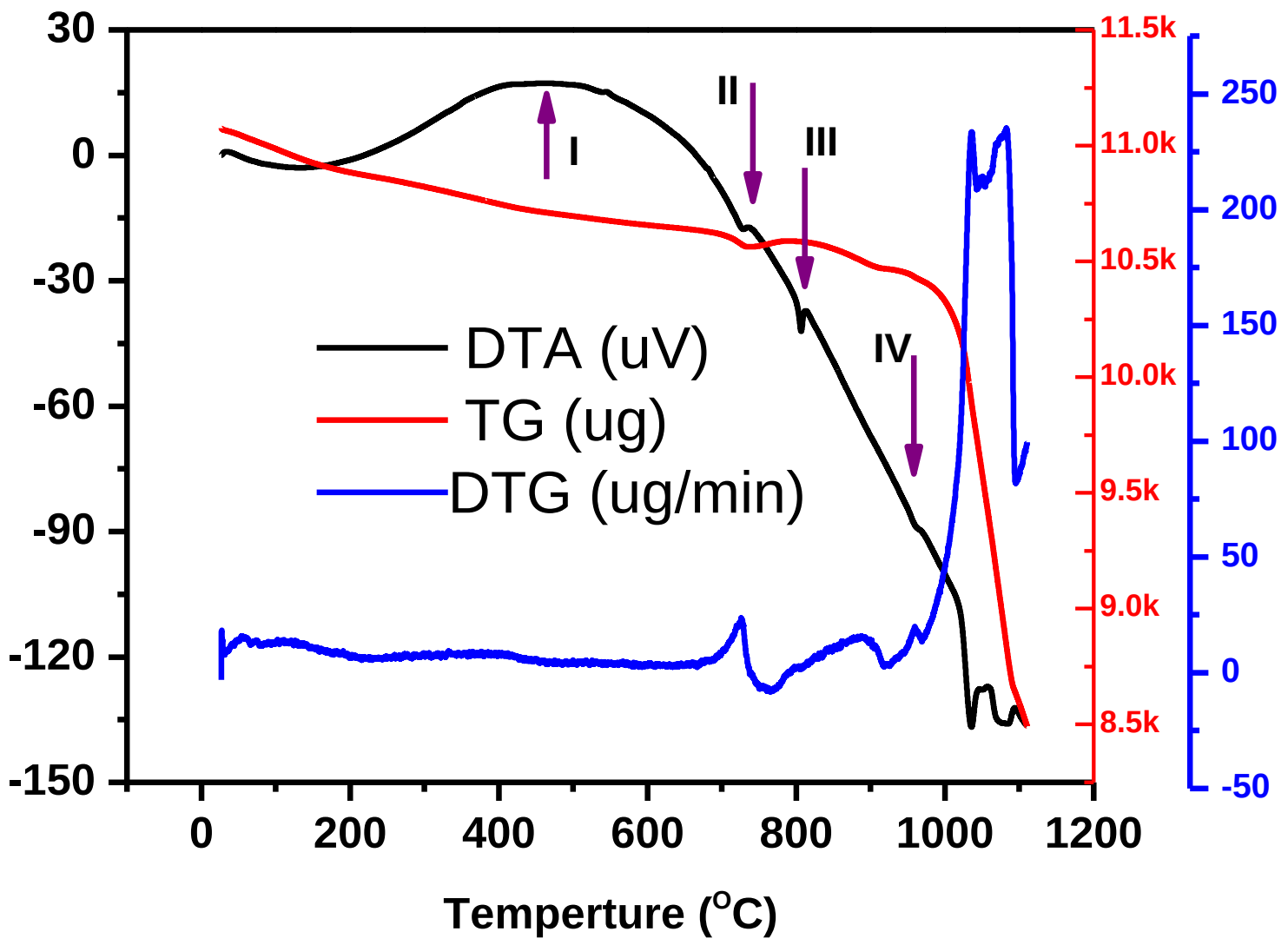

Fig. 1: DTA/TG/DTG curves for $Co_3O_4$ + $TeO_2$ mixture.

The thermo-gravimetric curves are divided in to four regions as marked I, II, III and IV in Fig.1. The region I, between 300 $^o$C to 600 $^o$C, corresponds to weight gain during the calcinations process. This weight gain can be seen as pre-requisite of "$O_2$" in the chemical reaction. The weight gain, as depicted by a hump in thermo-gravimetric curve is consistent with the earlier report of the CTO synthesis by Sisak and Luber [10]. The region II around 700 $^o$C and the region III around 800 $^o$C, are due to initiation of an intermediate phase formation and its transformation into the CTO phase, respectively. In the last region IV above 900 $^o$C, decomposition of the CTO phase into CoO and $TeO_2$ phase has been reported [10].

The mixed powder is calcined at different temperatures ranging from 500 $^o$C to 800 $^o$C for a fixed time of 12 hrs. Different calcined powders are designated here as CTO500, CTO600,

CTO650, CTO700 and CTO800. Fig. 2 compares SXRD patterns of the CTO500 and CTO600 calcined powder.

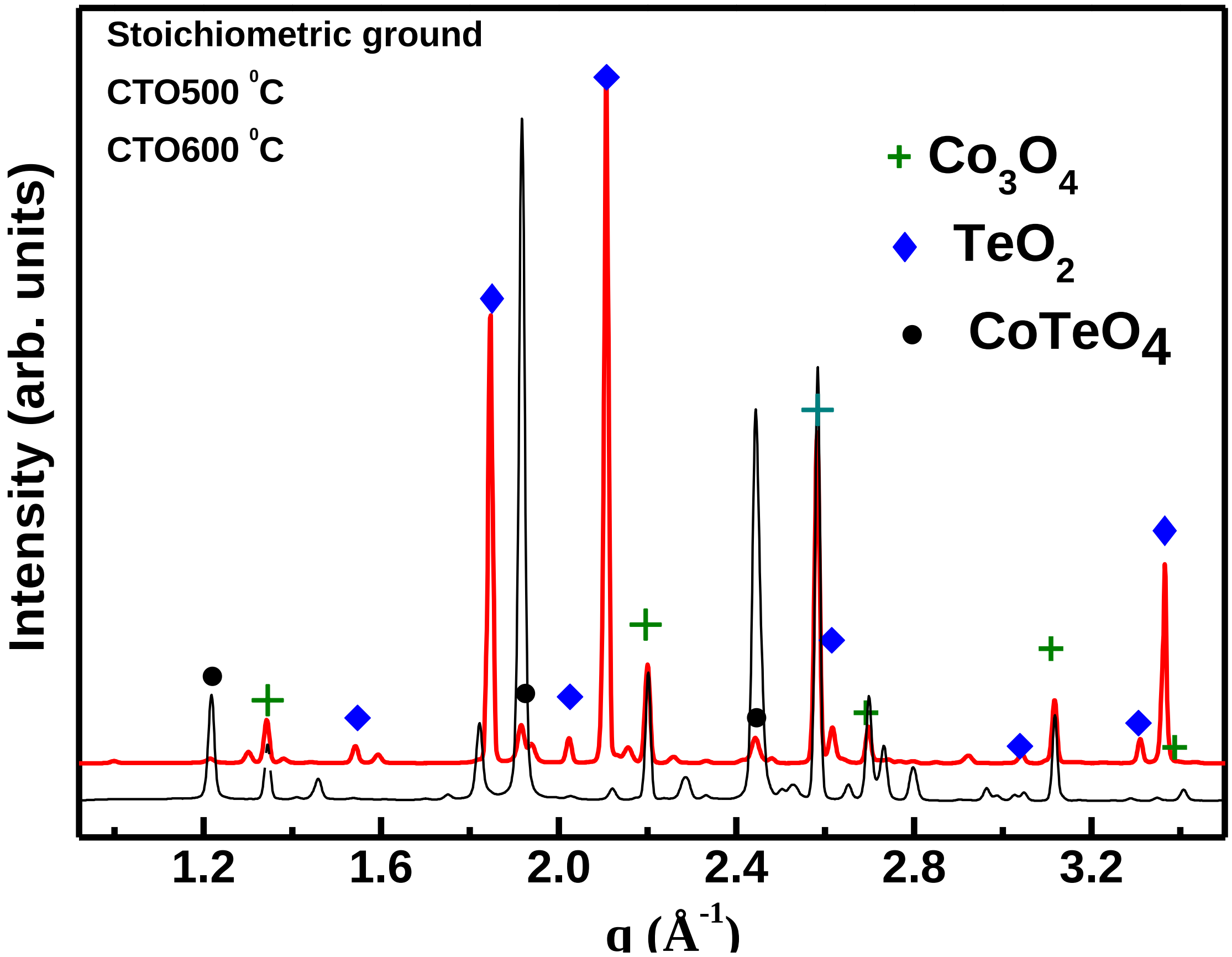

Fig. 2. The stoichiometric ground mixture at 500 $^0$C indicates initialization of a new phase ($CoTeO_4$) along with unreacted starting reactants (see text for details). Calcination at 600 $^0$C, on the other hand, corresponds to $CoTeO_4$ as major phase.

Initiation of a new phase formation, as marked by symbol "●", has been noticed for the CTO500 sample, which becomes major phase, when the mixed powder is calcined at 600 $^o$C. All other extra peaks have been assigned to un-reacted starting reagents $Co_3O_4$ and $TeO_2$. Rietveld refinement analysis of the CTO600 powder reveals formation of monoclinic $CoTeO_4$ ($P2_1/c$: $a$~ 6.192(3), $b$~ 4.671(3), $c$~5.567(2), $\beta$~124.07(2)$^0$) as an intermediate phase [20]. Along with the $CoTeO_4$ intermediate phase, commencement of the CTO phase has also been noticed in the CTO600 sample. Concentration of the CTO phase has been found to increase with further increase in the calcination temperatures. Fig. 3 compares the SXRD patterns of the CTO650, CTO700 and CTO800 samples. Phase analysis reveals that concentration of the CTO phase is increased along with equivalent decrease in the $CoTeO_4$ phase. No peak corresponds to the $CoTeO_4$ phase has been noticed in the CTO800 calcined powder. Although, few extra peaks corresponding to the $Co_3O_4$ has also been found in the SXRD pattern of the CTO800 sample, which is due to loss of the $TeO_2$ oxide consistent with its melting point of 790 $^{o}$C. Based on thermal and phase analysis, following reaction mechanism is suggested for the synthesis of the $CoTeO_4$ and CTO phase in the temperature regime 500 $^{o}$C to 900 $^{o}$C.

$Co_3O_4 + 3\ TeO_2 + O_2 \rightarrow 3\ CoTeO_4$--------(500 $^{0}$C-600 $^{0}$C) -------------------------------(1)

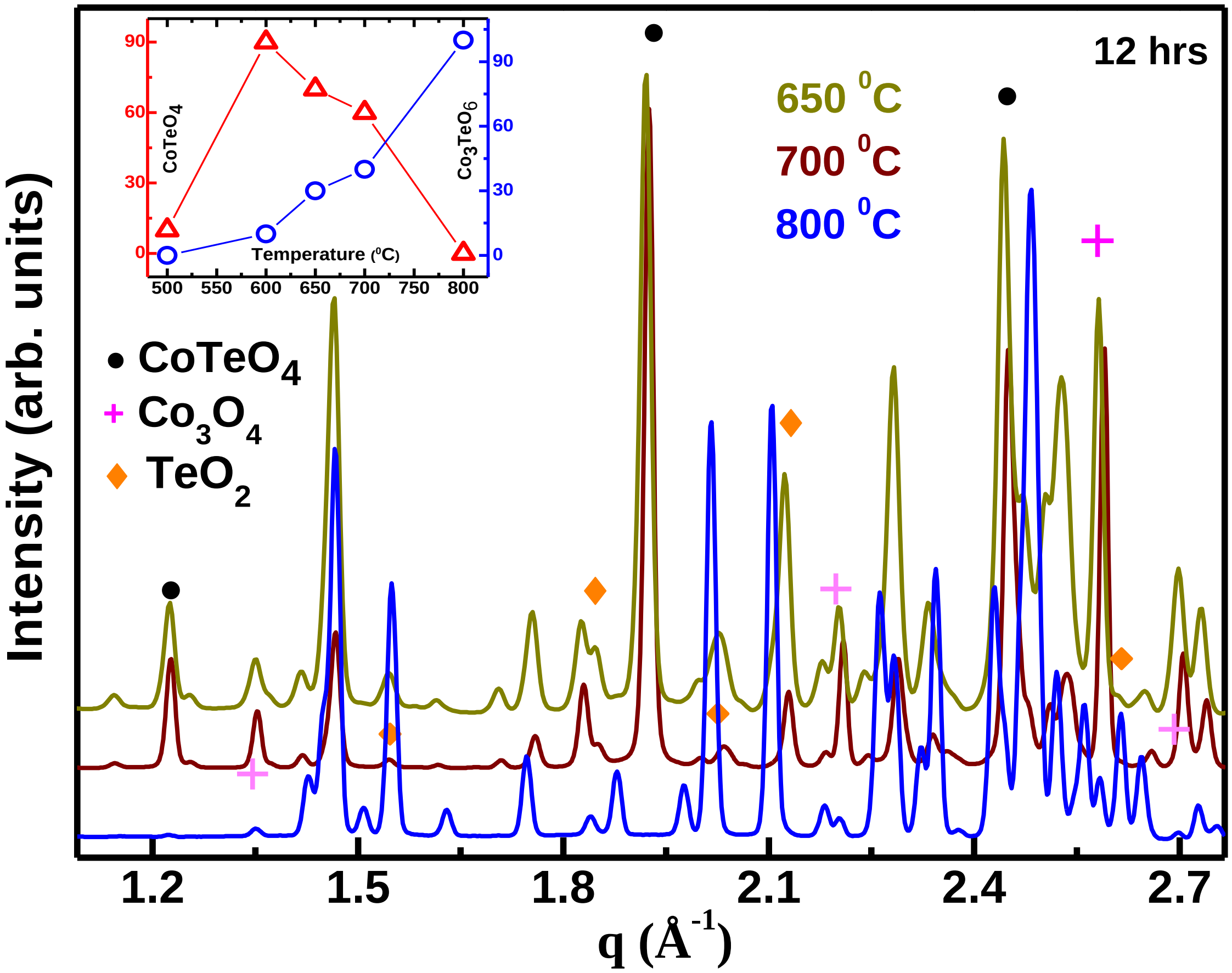

Fig. 3. Stabilization of $CoTeO_4$ along with various other phases, as we increase the calcinations temperature from 600 $^{0}$C to 800 $^{0}$C, all these phases are get converted to the final compound $Co_3TeO_6$ (see text for details).

This reaction mechanism is consistent with the weight gain observed in the thermal analysis plot shown in the Fig. 1 (region I).

$3\ CoTeO_4 + Co_3O_4 \rightarrow 2\ Co_3TeO_6 + TeO_2 + O_2$----(600 $^{0}$C -650 $^{0}$C)----(2)

$3\ CoTeO_4 + 2\ Co_3O_4 \rightarrow 3\ Co_3TeO_6 + O_2$ --------(650 $^0$C -700 $^0$C)-------------------------(3)

$3\ CoTeO_4 + 3\ Co_3O_4 + TeO_2 \rightarrow 4\ Co_3TeO_6 + O_2$ --------(700 $^0$C -800 $^0$C)----------------(4)

$2\ Co_3TeO_6 \rightarrow 6\ CoO + 2\ TeO_2 (\uparrow) + O_2$ --- (> 800 $^0$C)--(5)

Inset of the Fig. 1 shows variation of concentration of the $CoTeO_4$ and $Co_3TeO_6$ phases against the calcination temperature. Equation 6 has been used to calculate the concentration of $CoTeO_4/Co_3TeO_6$ phases.

$$\%\ \text{of}\ CoTeO_4/Co_3TeO_6 = I_{CoTeO4/Co3TeO6} / (I_{CoTeO_4} + I_{Co_3TeO_6}) \times 100 \text{ ------(6)}$$

It has been earlier reported that $TeO_2$ oxidizes to $TeO_3$, $Te_2O_5$ or $Te_4O_9$ around ~ 650 $^0$C -700 $^0$C. However, no such oxidation of the $TeO_2$ has been observed in the present study. Similarly no reduction of the $Co_3O_4$ into CoO and $Co_2O_3$ has been observed when the starting reagent $Co_3O_4$ is heated to 700 $^o$C. Phase analysis (SXRD not shown) of the $TeO_2$ and $Co_3O_4$ powder annealed at 700 $^0$C for 2 hrs, has confirmed the stability against oxidation or reduction of the starting reagents. All these above analyses are based on the single step synthesis in which the mixed ground powder is heated directly to all the aforementioned temperatures. Due to the large amount of unreacted $Co_3O_4$ phase in the last stage i.e. 800 $^o$C, double step synthesis has also been attempted in order to avoid the same. Fig. 4 compares SXRD pattern of CTO formation in a single step against double step synthesis in which the mixed powder is first heated at 700 $^o$C for 10-12 hours and then again at 800 $^o$C for 20-24 hrs with intermittent mixing. One can clearly see that lesser $Co_3O_4$ phase is present in the double step synthesis compared to that for the single step CTO synthesis. To get a clue about how much $Co_3O_4$ phase is present in both types CTO synthesis, percentage of the $Co_3O_4$ and CTO phase (with a total of 100) is calculated from the maximum peak intensity ratio through the SXRD patterns. Percentage of $Co_3O_4$ phase is calculated using equation

$$\%\ \text{of}\ Co_3O_4\ \text{phase} = [I_{Co_3O_4} / (I_{Co_3TeO_6} + I_{Co_3O_4})] \times 100, \text{------------(7)}$$

where $I_{Co_3O_4}$ and $I_{Co_3TeO_6}$ represent the highest peak intensity corresponding to the $Co_3O_4$ and CTO phase respectively. Above equation reveals a concentration of ~ 15-18 % and ~ 8-10 % $Co_3O_4$ phase for single and double step synthesis, respectively. It is important to note that the calcination temperature (800 $^{o}C$) is in between the melting point of $TeO_2$ (735 $^{o}C$) and $Co_3O_4$ (890 $^{o}C$), and presence of the $Co_3O_4$ is believed due to the $TeO_2$ loss during calcinations. It is not suitable to calcine the mixed powder in closed crucible for controlling the $TeO_2$ loss, thus extra amount of the $TeO_2$ has been added in the stoichiometric amount of $Co_3O_4$ and $TeO_2$ powders. However, several reports have pointed out the ambiguity in phase concentrations (due to different atomic numbers and symmetry of phases concern) calculated using the equation 7.

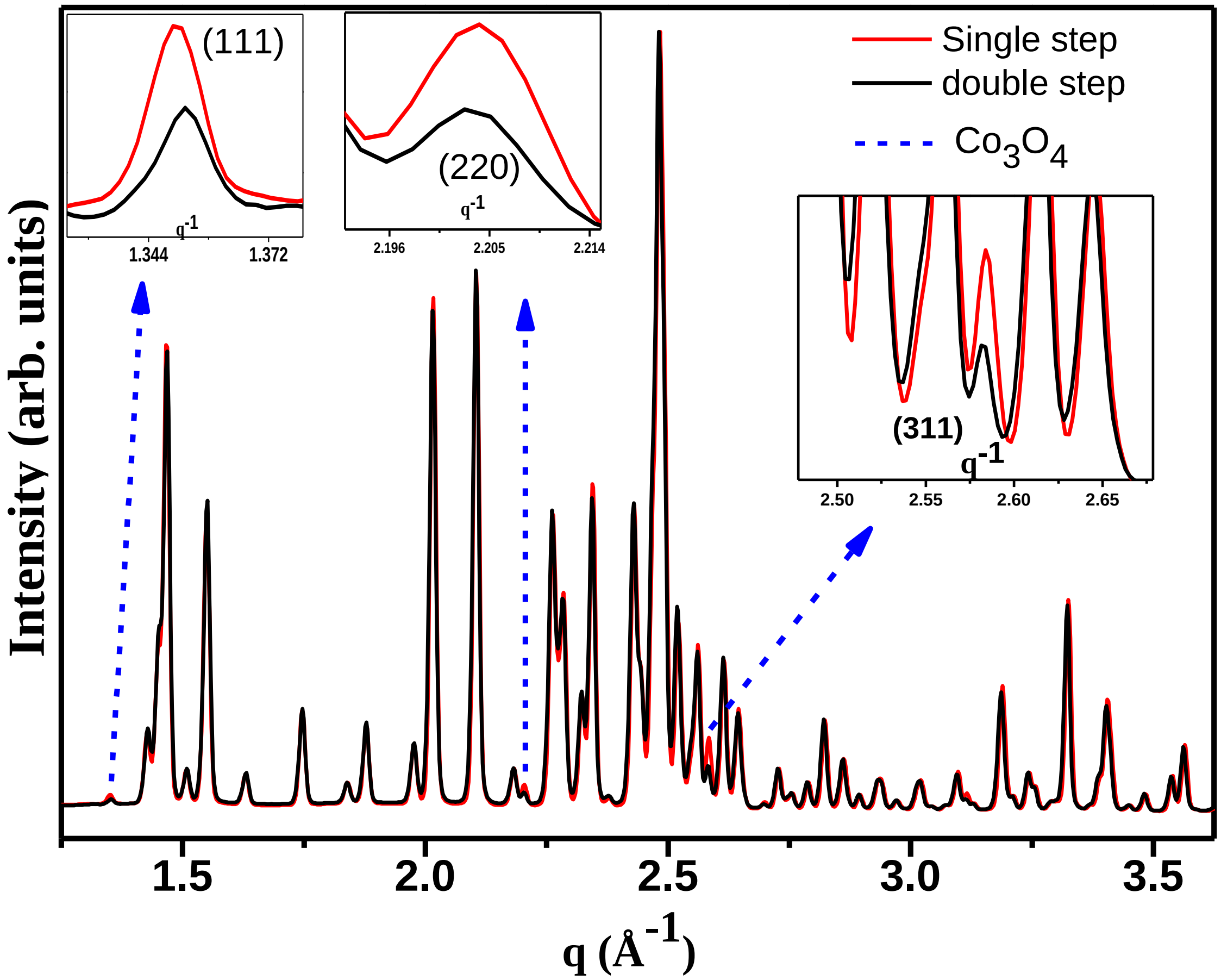

Fig.4. A) Shows comparisons of single step and double step calcinations while B) corresponds to stoichiometric vs non-stoichiometric calcinations (see text for details).

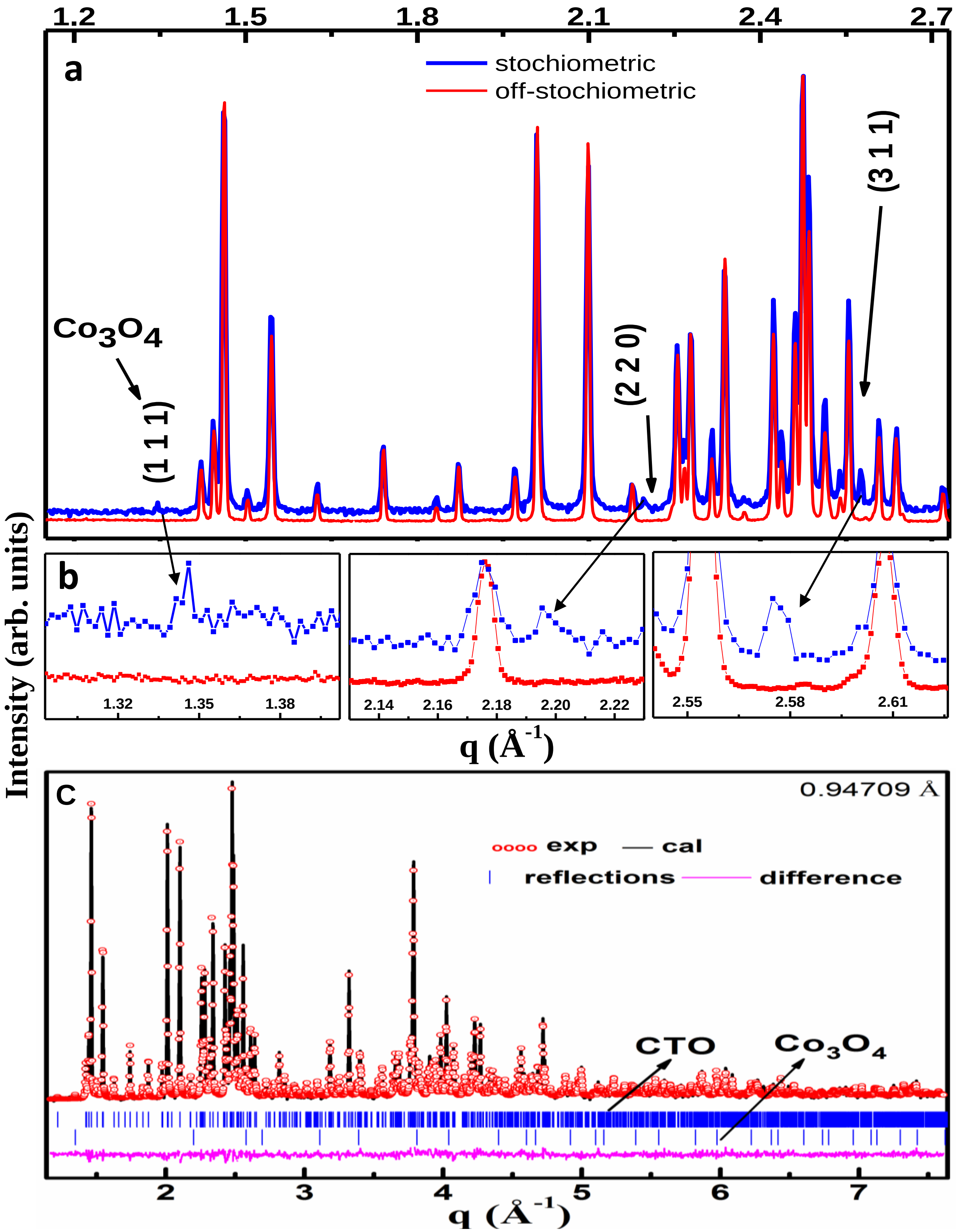

Fig. 5. a) A high resolution Synchrotron X-ray Diffraction measurement is used to probe monophasic CTO, b) show the corresponding impurity which is $Co_3O_4$, c) Rietveld refinement is used to estimate concentration of impure $Co_3O_4$ phase.

Therefore, in order to know the accurate value of the $Co_3O_4$ phase concentration, Rietveld refinement on the SXRD data has been used. For that, Fig. 5a shows high-resolution (HR) SXRD patterns for stoichiometric (blue color) and non- stoichiometric (red color) synthesis, carried out in the 2θ range of 10-70° recorded at room temperature. One can clearly see that $Co_3O_4$ phase is present in stoichiometric patterns with reference to the non-stoichiometric one, as shown in Fig. 5b (zoom part). Fig. 5 c shows two phases Rietveld Refinement on SXRD patterns along with experimental, calculated, and difference powder-diffraction profiles. The same quantitative phase composition analysis using Rietveld Refinement on SXRD patterns gives ~3-5 % $Co_3O_4$ phase in case of the double step CTO synthesis. To balance this loss in $TeO_2$, extra $TeO_2$ has been added to complete the reaction. Fig. 5 compares the SXRD patterns of stoichiometric (without extra $TeO_2$) and non-stoichiometric (with extra $TeO_2$) CTO synthesis. Thus, single phase CTO powder is synthesized when ~ 5 wt.% $TeO_2$ extra amount has been added in the stoichiometric amount of $Co_3O_4$ and $TeO_2$ and the mixed powder is first calcined in air at 700 $^{o}$C for 10 hrs and then again at 800 $^{o}$C for 24 hrs with intermittent mixing and drying steps. The above process has been repeated number of times to confirm the single phase CTO formation.

Detailed Rietveld refinements has been done for the insight into the structural investigation of pure CTO and the results are in a good agreement with earlier crushed single crystal as well as powder XRD data [14,15]. Regarding insight into the structure of CTO, which contains five Co, two Te and nine O, Rietveld refinements and the corresponding valence bond sum calculations has also been done on SXRD data. Crystallographically distinct Co cations occupy 4e and 8f Wyckoff positions, Te occupies 4b and 8f and all O occupies at 8f sites. Both the Te exhibits octahedral sites and five Co sits both in tetra and octahedral sites. In order to verify the structural aspects of monophasic ceramic CTO, thorough Rietveld refinements and their structural parameters on Synchrotron X-ray Diffraction data can be found elsewhere [5,6,21].

As discussed in part of introduction, origin of multiple magnetic transitions in single phasic CTO remains unresolved. For example, whether one of the transitions occurred at ~34 K belongs to intrinsic CTO or some impure phase. In this reference, focus has been drawn towards $Co_3O_4$ which shows AFM transition ~ 33 K [22]. Fig. 6 shows DC magnetization curve of ceramic CTO. Main panel of Fig. 6 (and its left inset), which corresponds to stoichiometric CTO (which contains $Co_3O_4$ impurity), shows a transition at around 34 K (shown by arrow) consistent with that of single crystal as well as our ceramic CTO [4,6,23]. Remaining magnetic transitions such as 26 K and 18 K is consistent with the AFM transitions reported for CTO [4,6,23].

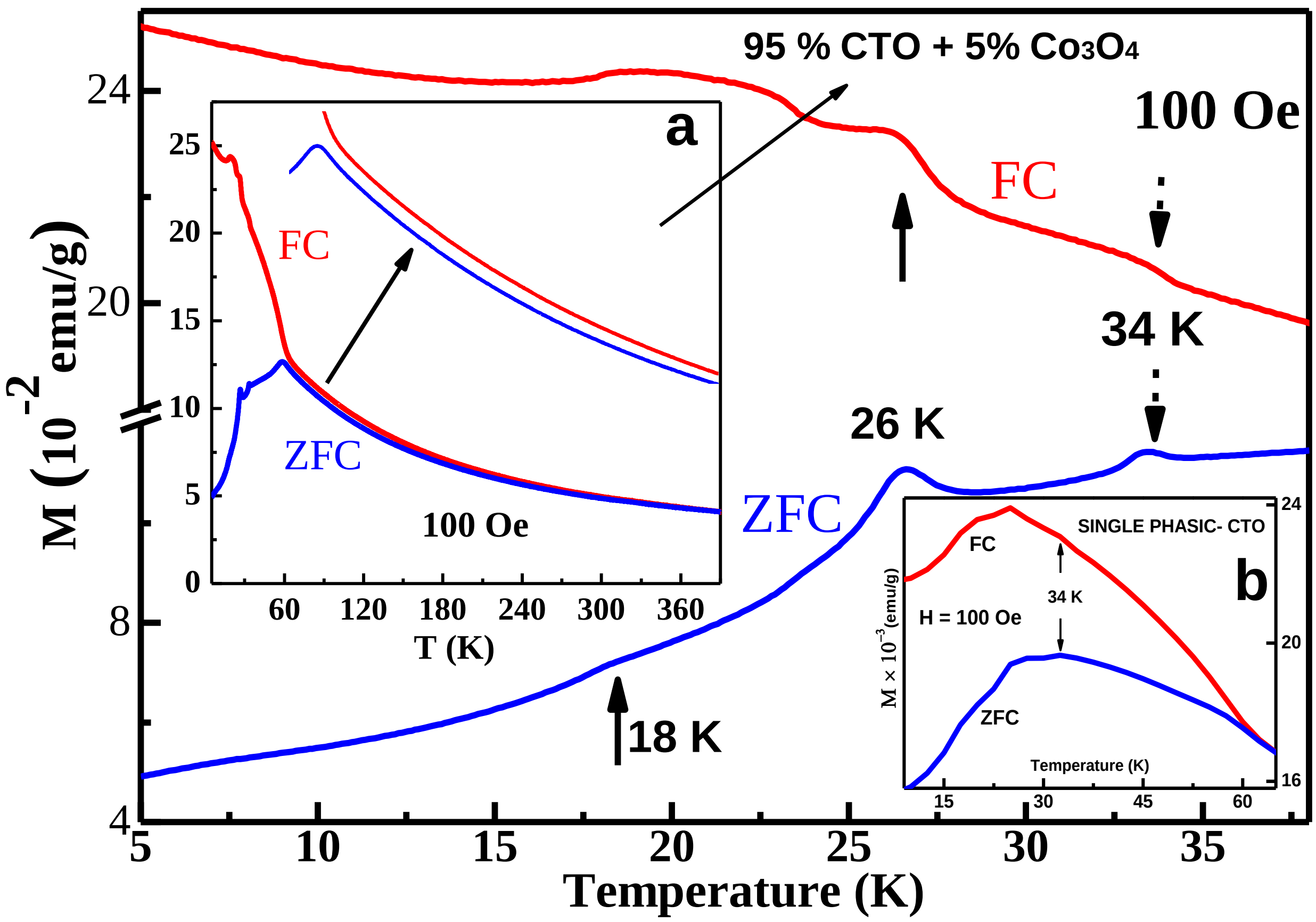

Fig. 6. Main panel as well as inset (a) shows DC magnetization data for impure (CTO + $Co_3O_4$) sample showing effect of $Co_3O_4$ on CTO magnetization data. Inset (a) shows separation between ZFC and FC curves of the same data above 60 K, and inset (b) shows the same for mono-phasic CTO (see text for more details).

The right inset of Fig. 6 (shown for clarity) shows the same for non-stoichiometric CTO (which does not contains any impurity as per our HR-SXRD). It also shows all the aforementioned magnetic transitions as observed in stoichiometric CTO. Analysis of magnetic data indicates that this ~34 K transition is intrinsic to CTO, not because of $Co_3O_4$ phase. The only effect of impure $Co_3O_4$ phase at CTO magnetization data is the nature of zero field cooled (ZFC) and field cooled (FC) curves above 60 K, and saturation like behavior of FC below 25 K. Interestingly, our magnetization data (both for impure and pure CTO) shows a bifurcation in ZFC/FC at around 60 K, which is not the case for single crystal, indicates additional intrinsic nature of our ceramic CTO. Further attention has been drawn towards the nature of cations (Co and Te ions) valency in order to guess additional magnetism in our ceramic CTO.

For the same purpose, XANES measurements have been performed. Fig. 7 shows the XANES spectra of Te L-edge (main panel) and Co K edge (inset) along with their individual standards. Details and the corresponding data analysis of a typical XANES spectrum can be found elsewhere [6,24,25]. Following these, comparison of Co K-edge XANES spectra with their standards indicates the mix valency of Co in the single phasic ceramic CTO. Whereas earlier reports on the CTO claim (without any experimental evidence) that Co in CTO is to be in $Co^{2+}$ [4,7,14,15,23]. On the other hand, comparison of Te L- edge XANES spectra with their standards clearly indicates $Te^{6+}$ state in CTO. Due to this unambiguous results obtained from XANES analysis, XPS measurements has also performed.

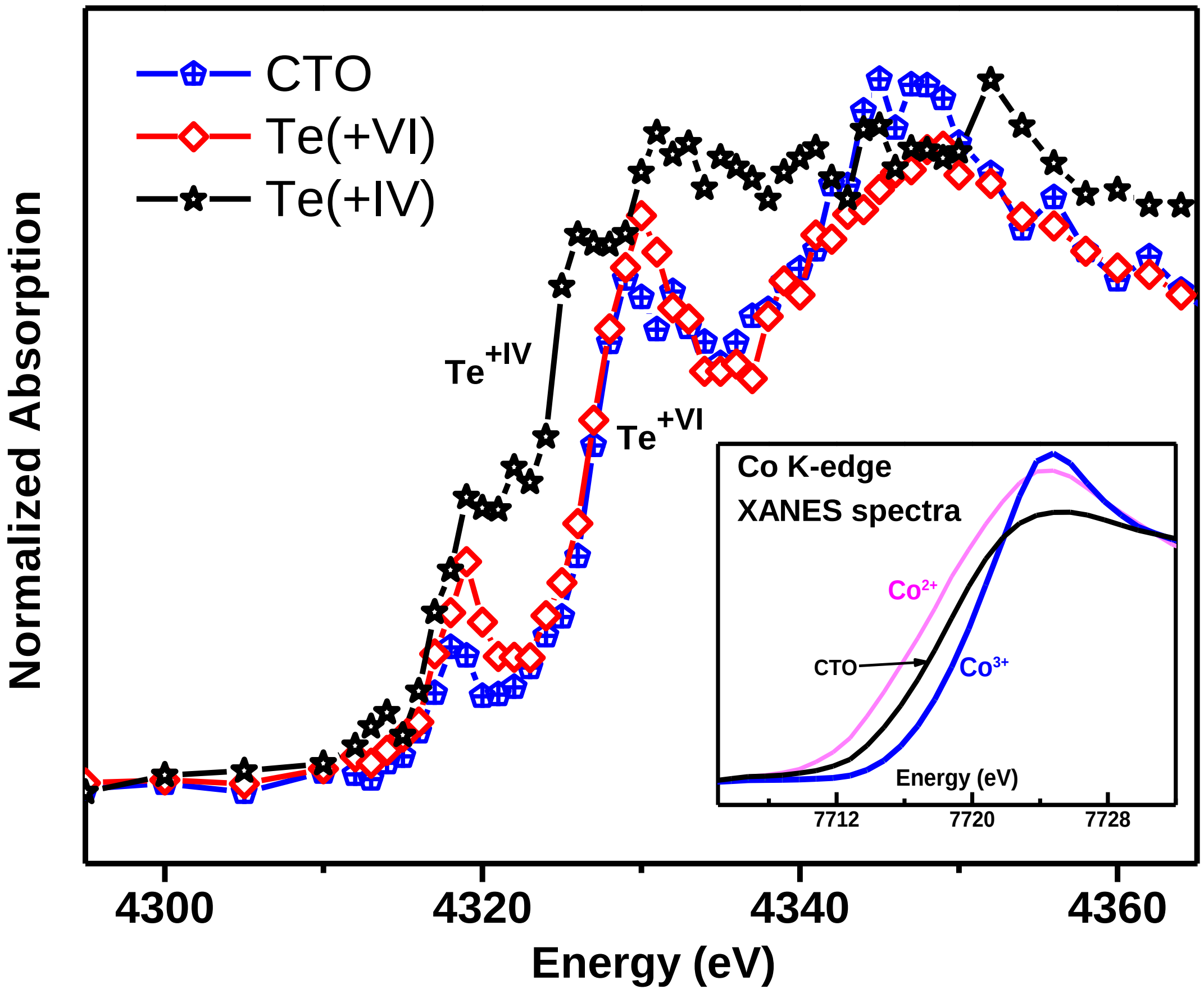

Fig. 7: Te $L_3$ edge XANES spectra of CTO along with Te standards $Te^{+4}$ and $Te^{+6}$. Inset shows Co-K-edge XANES spectra of CTO with $Co^{2+}$ and $Co^{3+}$.

To cross check the XANES results about the Co and Te valence, the X-ray photoelectron spectroscopic measurements are performed. Fig. 8 shows XPS spectra of prepared CTO samples for Co 2p, Te 3d and O 1s edges. The photoelectron spectral features of oxygen (O1s) peak for the sample falls at 529.93 ± 0.1eV (I: oxide peak and II: adsorbed hydroxyl group), [25] clearly showing the purity of $Co_3TeO_6$. This is in agreement with the XANES as well HR-SXRD results. The position of O 1s peak matches well with the tabulated value of 529.9 eV for $Co^{2+}/Co^{3+}$ based compound [25,26,27]. Co 2p spectra for our CTO sample shows main Co $2p^{1/2}$ and Co $2p^{3/2}$ peaks at 780.05 eV and 795.80 eV respectively, with significant satellite peaks. By these main peak (binding energy) positions only, one can't probe strongly, the exact valence state of Co in the samples [27]. It is only the satellite peak intensity along with their binding energy positions that identify the Co valence state.

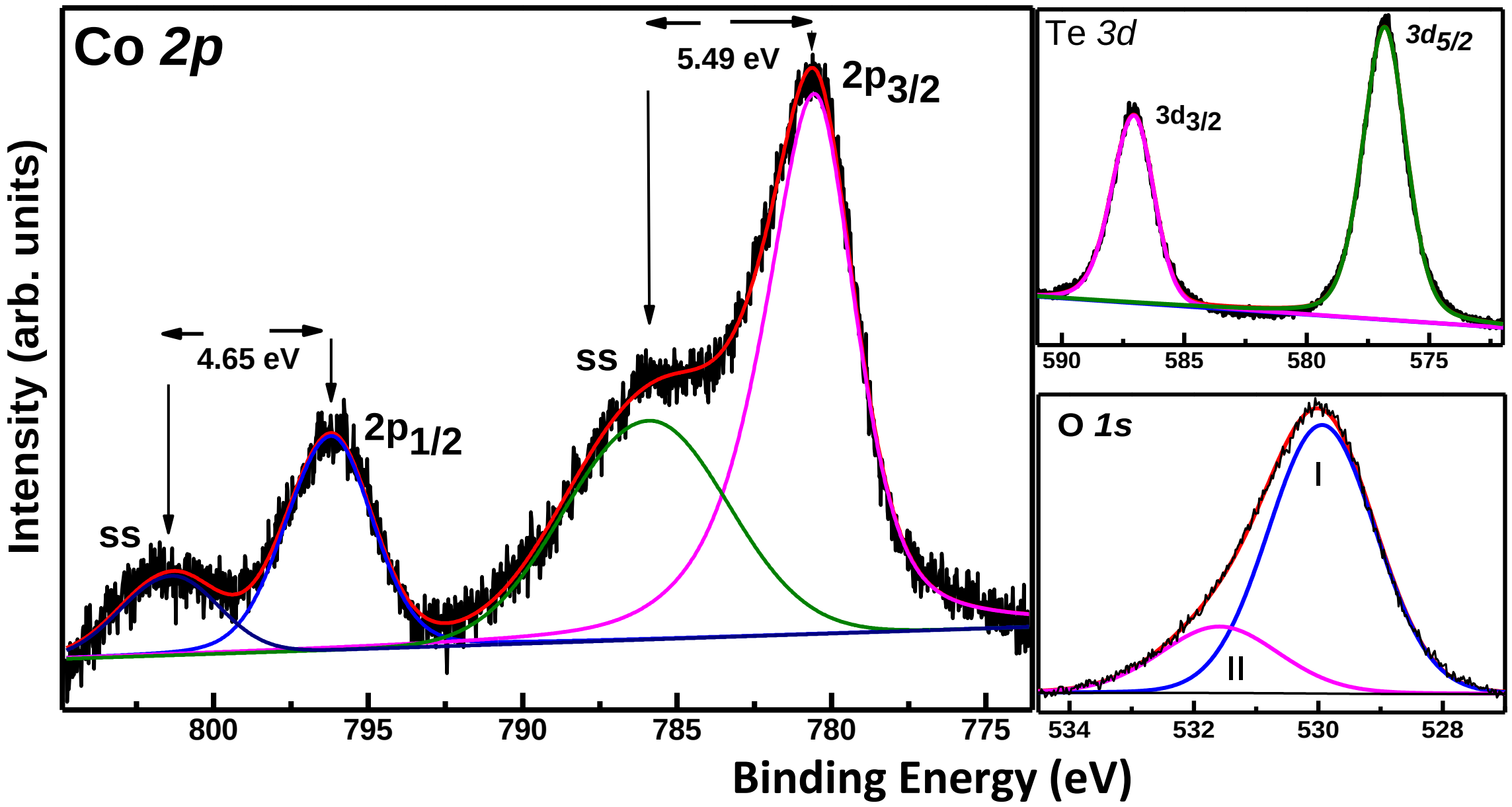

Fig. 8. a) Normalized core level XPS spectra of Co *2p,* satellite (ss) peaks nearby main peaks confirms the mixed oxidation state of Co in CTO, b) normalized core level XPS spectra of Te *3d*, *b*inding energy 587.3 eV and 576.8 eV positions of the main peaks confirm the +VI oxidation state of Te and c) Oxygen *1s* XPS spectra CTO sample fitted with Shirley background and two Lorentzian-Gaussian peaks acquired with Al Kα source.

Table I. Binding energy positions (in eV) for Co 2*p* which represent Co (II)/Co(III) states are the most prominent states in our CTO. O/T denotes the octahedral and tetrahedral coordination respectively. While, the same for Te 3*d* represent the + 6 state of Te in CTO.

| | | | Co 2*p* | | |
|---|---|---|---|---|---|
| **Sr. No.** | **$2p_{3/2}$** | **ss** | **$2p_{1/2}$** | **ss** | **Ref (No.)** |
| **1- $Co^{2+}$(O/T)** | **780.5/780.7** | **786.4/789.5** | **796.3/796.0** | **803.0/804.5** | **[22,27]** |
| **$Co^{3+}$(O)** | **779.6** | **---** | **794.5** | **----** | **[22,27]** |
| **2** | **780.05** | **784.70** | **795.80** | **801.29** | **CTO** |
| | | | **Te 3*d*** | | |
| | **$3d_{5/2}$** | | **$3d_{3/2}$** | | |
| **1** | **576.8** | **--** | **587.2** | **--** | **[28]** |
| **2** | **576.6** | **--** | **587.0** | **--** | **[29]** |
| **3** | **576.87** | **--** | **587.18** | **--** | **CTO** |

The satellite peak for Co $2p^{1/2}$ and Co $2p^{3/2}$ peaks are shifted by (4.65 ± 0.1) eV and (5.49 ± 0.1) eV, respectively from their main peaks. As only two ionic state of Co (+2 & +3) are stable, in the present scenario, so a comparative table for these two is shown in Table I. From the table one can clearly see the mix valence behaviour of Co in the pure samples. The satellite peaks are attributed to crystal field splitting for $Co^{2+}$ in tetrahedral crystal field environment. The separation between Co $2p^{1/2}$ and Co 2 $p^{3/2}$ main peaks lie within (15.75 ± 0.1) eV for CTO sample. However, spin-orbit multiplet separation is not sensitive to be used for identification of ionic valence state in cobalt oxide [25].

Further, the Te 3d peak position (FWHM) has been found at 576.8 (1.96) eV for $3d^{5/2}$ and 587.2 (2.08) eV for $3d^{3/2}$, with spin orbit separation of 10.2 eV. This peak corresponds to Te 3d peaks of Te (VI) state with no satellite peaks as usual for Te (VI) state [28,29]. Table I compares for Te (+VI) with reported states of Te. XPS analysis so far confirms the mixed valence Co and VI$^{+}$ oxidation states of Te in the CTO sample, in agreement with XANES results.

Combined XANES and XPS analysis on our ceramic CTO clearly indicate various possibilities such as oxygen non-stoichiometry [30], or cations non-stoichiometry [31,32]. For example, observed mix valency of Co and +VI valency for Te suggest oxygen non-stoichiometry in CTO. Existence of mixed valence Co in CTO is likely to be the case of Co in spinel $Co_3O_4$ as per earlier discussion. The most likely cause for charge dis-balance in an oxide is O non-stoichiometry. However, the other cause for the same is cations excess or vacancy, which is not an easy task to identify the same at least for CTO. On comparing the binding energy positions of Co, Te and O peaks from the aforementioned table, and also from XANES, one may have clear clue on the origin behind the induction of $Co^{3+}$ in CTO in addition to $Co^{2+}$.

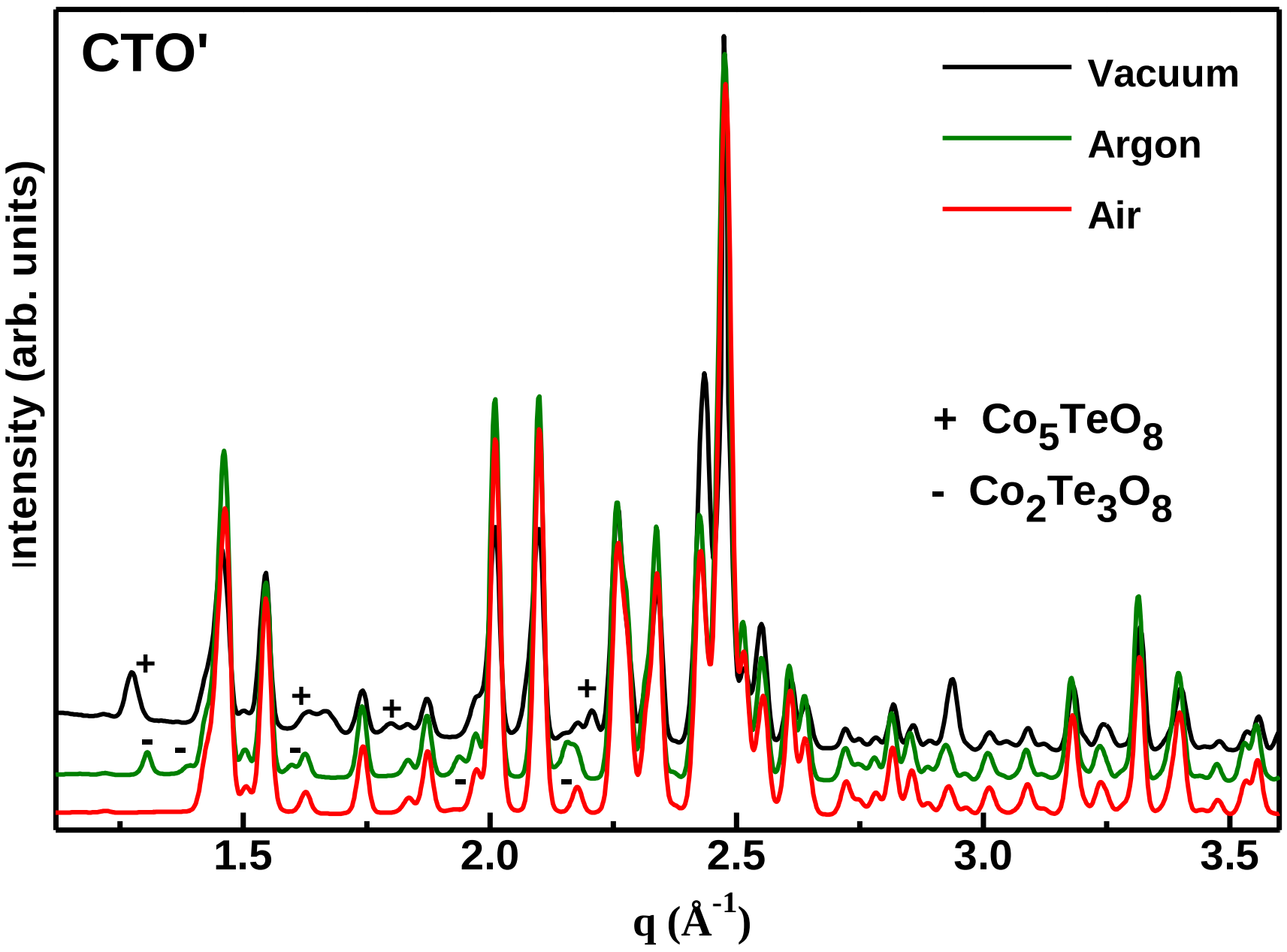

Fig.9: SXRD patterns of CTO synthesized in air, argon and vacuum atmosphere, indicating the only novel route for CTO synthesis.

In the above discussion, the origin of $Co^{3+}$ in the ceramic CTO seems to be due to oxygen non-stoichiometry. To confirm this, the same synthesis was carried out in different (oxygen reduced) atmospheric conditions. In order to grow CTO in different atmospheres, we have chosen argon as well as vacuum as oxygen reduced atmospheres. In the argon, tube furnace was evacuated up to $2\times10^{-3}$ mbar pressure and applied the temperature as earlier i.e. two step calcinations. Whereas in the case of vacuum calcination, pressed pellet of ground mixture was sealed in a quartz ampoule at $7\times10^{-6}$ mbar pressure. Then the ampoule is put into the tube furnace for the required calcinations. Obtained sample is then characterized by aforementioned techniques. Fig. 9 shows SXRD patterns for the CTO samples prepared under different atmosphere. This indicates that only the CTO synthesis in air atmosphere results in monophasic CTO. Rest of the CTO samples is impure with an impurity either $Co_5TeO_8$ (JCPDS # 20-0367) or $Co_2Te_3O_8$ (JCPDS # 89-4451).

We also mention the possible reaction below.

$$(2/3)\ Co_3O_4 + (3)\ TeO_2 \rightarrow Co_2Te_3O_8 + (2/3)\ O_2\ (g) \text{ ----Argon----}(650\ ^0C\text{-}800\ ^0C)\text{---}(9)$$

$$(5/3)\ Co_3O_4 + TeO_2 \rightarrow Co_5TeO_8 + (2/3)\ O_2\ (g) \text{ ----Vacuum---- }(650\ ^0C\text{-}800\ ^0C)\text{----- }(10)$$

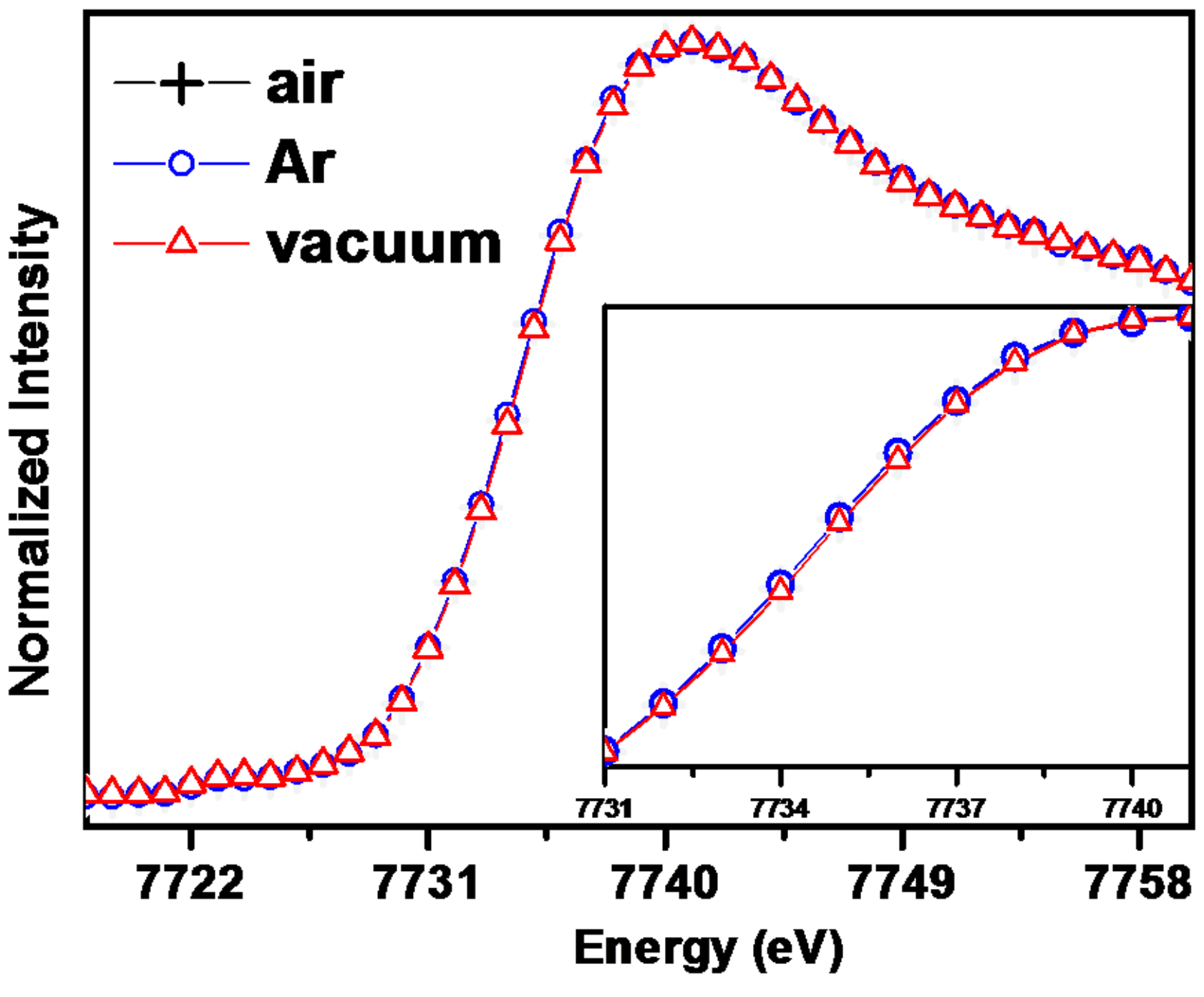

Fig. 10: Co K edge XANES spectra of CTO synthesized in air, argon and vacuum atmosphere, indicates same valency.

Another important observation from these samples is the same valency of Co as found in earlier case i.e. air calcined sample. This observation has been carried out using XANES measurements at Co K edge, where XANES spectra show same oxidation state of Co in all the samples. Fig. 10 shows the edge step normalized XANES spectra at Co K edge for all the samples prepared in different conditions. This actually complicates the cause of $Co^{3+}$ in CTO. Based on all these observation, we can conclude that single phase of CTO forms only in air with two step calcinations. Combined analysis using aforementioned techniques indicates possibility of cations vacancy in our prepared samples.

To summarize the work, ceramic $Co_3TeO_6$ is synthesized via conventional solid state reaction route and possible reaction mechanism has been proposed. Only off-stoichiometric $Co_3O_4$ and $TeO_2$ and two steps calcination in air leads to single phase CTO. CTO structure crystallizes in monoclinic C*2/c* space group in agreement with the earlier structure reports of single crystal as well as powder CTO. The existence of $Co^{3+}$ ions does not diminish any of the multiferroic property seen earlier in powder as well as in single crystal CTO, but enhances some of the observables [5,6]. For example, appearance of $Co^{3+}$ ions in $Co^{2+}$--O--$Co^{2+}$ network can cause (a) structural distortion, (b) local FM interaction through double-exchange, (c) weakening of effective magnetic interaction between the $Co^{2+}$ ions and (d) in-commensurate (IC) AFM ordering. For instance, a possible reason for appearance of the IC-AFM order in $Co_3O_4$ is considered to be local structural modulations due to change in charge/spin state of Co ions [22]. The magnetic behavior is similar to that of single crystal CTO at high field, but exhibits slightly different observations at very lower magnetic fields [5,6]. Here, on the other hand, the insight into the growth of ceramic CTO has been discussed.

## 4. Conclusions

Two step solid state reactions route and possible growth reaction mechanism for the synthesis of single phasic cobalt tellurate $Co_3TeO_6$ (CTO) using $Co_3O_4$ and $TeO_2$ has been proposed. Synthesis optimization has been done using High Resolution Synchrotron X-ray Diffraction (HR-SXRD). An intermediate $CoTeO_4$ phase has been shown to transform directly into CTO leading to single phase. HR-SXRD reveals that out of the stoichiometric and off-stoichiometric synthesis, only off-stoichiometric *mixture* results *monophasic* CTO. Further, core level study using XANES and XPS indicate multiple oxidation states of Co in the CTO. DC-magnetization, which shows a transition at 34 K apart from two transitions at 26 K and 18 K, is intrinsic to CTO and not due to $Co_3O_4$. Moreover, enhanced multiferroic properties such as effective magnetic moment have been correlated with the present synthesis route.

**Acknowledgements**

Authors are grateful to Dr. P. A. Naik, Dr. G. S. Lodha and Dr. P. D. Gupta for their support and encouragement. We thank Dr. A. Sundaresan, B. Rajeswaran for initial magnetic measurements. We thank Dr. S. N. Jha, Dr. P. Rajput and Ashok K. Yadav for Te L edge XANES measurements. We also thank Dr. D. M. Phase and A. D. Vadikar for performing XPS measurements and Dr. Piyush K. Patel for taking thermo-gravimetric data. HS acknowledges Homi Bhabha National Institute, India for research fellowship.